\documentstyle[12pt,epsfig]{kluwer}
\begin{document}
\begin{article}

\begin{opening}
\title{Reversed Drifting Quasi-periodic Pulsating Structure in an
X1.3 Solar Flare on 2005 July 30}
\author{Rui \surname{Wang}}
\author{Baolin \surname{Tan}}
\author{Chengming \surname{Tan}}
\author{Yihua \surname{Yan}}
\institute{Key Laboratory of Solar Activity, National Astronomical
Observatories, \\Chinese Academy of Sciences, Beijing, 100012,
China. \\Email: Ray@nao.cas.cn}
\date{Received: ****; Accepted: 7 January 2012}
\begin{abstract}

Based on the analysis of the microwave observations at frequency of
2.60 -- 3.80 GHz in a solar X1.3 flare event observed at {\it Solar
Broadband RadioSpectrometer} in Huairou (SBRS/Huairou) on 2005 July
30, an interesting reversed drifting quasi-periodic pulsating
structure (R-DPS) is confirmed. The R-DPS is mainly composed of two
drifting pulsating components: one is a relatively slow very
short-period pulsation (VSP) with period of about 130 -- 170 ms, the
other is a relatively fast VSP with period of about 70 -- 80 ms. The
R-DPS has a weak left-handed circular polarization. Based on the
synthetic investigations of {\it Reuven Ramaty High Energy Solar
Spectroscopic Imaging} (RHESSI) hard X-ray, {\it Geostationary
Operational Environmental Satellite} (GOES) soft X-ray observation,
and magnetic field extrapolation, we suggest the R-DPS possibly
reflects flaring dynamic processes of the emission source regions.

\end{abstract}
\end{opening}

\section{Introduction}

Quasi-periodic pulsations (QPPs) associated with solar flares are
observed frequently in optical, EUV, soft X-ray, hard X-ray, and
radio emissions (see the recent review of Nakariakov and Melnikov,
2009). For pulsation events, Aschwanden (1987, 2004) presented an
extensive review about the models, and classified them mainly into
three groups: (1) magnetohydrodynamic (MHD) flux tube oscillations
(eigenmodes); (2) Cyclic self-organizing systems of plasma
instabilities; (3) Modulation of periodic electron acceleration.
Based on the radio observations and the period of pulsation (P),
QPPs can be classified into three types (Wang and Xie, 2000): (1)
long period pulsation (LPP), P $\sim$ tens of seconds; (2) short
period pulsation (SPP), P $\sim$ several seconds; and (3) very short
period pulsation (VSP), P $\sim$ subseconds. Recently, some
supplements and extensions had made the classification more
comprehensive and detailed (Tan et al, 2010). The very long period
pulsation (VLP) was added, whose period is in the hectosecond or
several minutes range. Generally it is defined as P $>$ 100 s. On
the other hand, the VSP was divided into two sub-classes: slow-VSP,
where the period is in the decisecond, 0.1 $<$ P $<$ 1.0 s and the
other is fast-VSP, P $<$ 0.1 s. The different QPPs should be
corresponding to different generation mechanisms, and might reveal
different physical conditions in the source region. The
flare-associated QPP can provide information of solar flaring
regions, and give some prospective insight into coronal plasma
dynamic processes, providing remote diagnostics of the microphysics
of energy release sites. The understanding of flaring QPP in the
solar corona will open up very interesting perspectives for the
diagnostics of stellar coronae (Mathioudakis, et al., 2003).

Usually, some QPPs frequently have another important feature:
frequency-time drift, recognized as drifting quasi-periodic
pulsating structures (DPS) (Kliem, Karlick\'y, and Benz, 2000). The
analysis of the frequency drift rate in DPS may provide information
not only about the dynamical processes of the source region but it
can also reveal atmospheric properties. Kliem, Karlick\'y, and Benz
(2000) proposed a model in which the decimetric DPS is caused by
quasi-periodic particle acceleration episodes that result from a
highly dynamic regime of magnetic reconnection in an extended
large-scale current sheet above the soft X-ray flare loop, where
reconnection is dominated by repeated formation and subsequent
coalescence of magnetic islands, known as secondary tearing modes.
With this model, they explained the global frequency drifting
pulsating structure as a motion of the plasmoid in the solar
atmosphere with density gradient. Here, particles are accelerated
near the magnetic X-points in the DC electric field associated with
magnetic reconnection. The strongest electric fields occur at the
main magnetic X-points adjacent to the plasmoid, and a large
fraction of the accelerated particles may be temporarily trapped in
the plasmoid; the accelerated process itself may form an anisotropic
velocity distribution, which excites the observed radio emission. In
fact, there are a series of works to explain DPSs as the radio
emission being generated during multi-scale tearing and coalescence
processes in the extended current sheet of a flare (Karlicky, 2004;
Karlicky et al., 2005). Based on particle-in-cell simulation,
Karlicky and Barta (2007) found that electrons are accelerated most
efficiently around the X-point of the magnetic configuration at the
end of the tearing process and the beginning of plasmoid
coalescence. The most energetic electrons are mainly localized along
the X-lines of the magnetic configuration.

However, so far, from the observations, we only obtained DPS with
single-directional frequency drifting rate, i.e., drift from high
frequency to lower frequency, or from low frequency to higher
frequency in the single DPS event. A DPS with double-directional
frequency drifting rate, i.e., the emission drifts from higher
frequency to lower and then reversed, namely from lower frequency
to higher may be called as reversed drifting quasi-periodic
pulsating structure (R-DPS). By scrutinizing the microwave
observation data obtained in {\it Chinese Solar Broadband
RadioSpectrometer} (SBRS/Huairou), we find a particular example of
R-DPS in the flare on 2005 July 30.

This paper is arranged as follows. Section 2 presents the
observational data and the data analysis. Section 3 gives some
discussions on physical processes related to the R-DPS. Finally,
Section 4 draws our conclusions.

\section{Observations and Data Analysis}

\subsection{Observations}

On 2005 July 30 an X1.3 flare/CME event occurred from 06:10 UT to
07:00 UT, with the peak at 06:35 UT in AR 10792 at N11$^\circ$,
E52$^\circ$, near the east edge of the solar disk. During this flare
event, several solar telescopes got the perfect observational data,
such as the solar microwave (SBRS/Huairou), {\it Reuven Ramaty High
Energy Solar Spectroscopic Imaging} (RHESSI) hard X-ray, {\it
Geostationary Operational Environmental Satellite} (GOES) soft
X-ray, optical {\it Michelson Doppler Imager} on {\it Solar and
Heliospheric Observatory }(MDI/SOHO), and {\it Big Bear Solar
Observatory} (BBSO), etc. In this work, our focus is on the
microwave observations. We mainly use the observation of
SBRS/Huairou to investigate the properties of QPP. SBRS/Huairou
includes three parts: 1.10 -- 2.06 GHz, 2.60 -- 3.80 GHZ and 5.20 --
7.60 GHz (Fu et al., 1995; Fu et al., 2004; Yan et al., 2002). R-DPS
appeared in the frequency band of 2.60 -- 3.80 GHz and the time
range from 06:24:15 -- 06:24:21 UT, and the duration lasts for about
6 s (Figure 1).

\begin{figure}
\begin{center}
 \includegraphics[width=10.0cm]{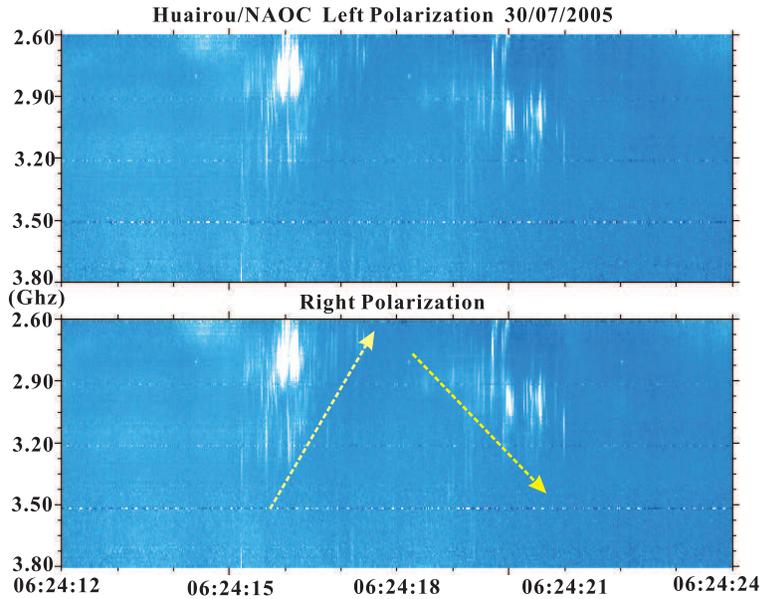}
  \caption{The spectrum of the solar microwave drifting quasi-periodic pulsating structure
  at 06:24:12 UT -- 06:24:24 UT, on 2005 July 30, observed at SBRS/Huairou with the spectrometer of
   2.60 -- 3.80 GHZ. Two yellow arrows indicate frequency drifting directions.}
\end{center}
\end{figure}

The antenna diameter of the SBRS/Huairou at frequency of 2.60 --
3.80 GHz is 3.2 m. It is controlled by a computer to automatically
trace the solar disk center and can receive the total flux of solar
radio emission with dual circular polarization. The dynamic range of
this instrument is 10dB above quiet solar background emission and
the observation sensitivity is $\triangle S/S_{\odot}\leq 2\%$,
where S$_\odot$ is the quiet solar background emission. The data
processing used the software in IDL language and data calibration
followed the method proposed by Tanaka et al. (1973). The standard
flux values of the quiet Sun are adopted from the data published by
the Solar Geophysical Data (SGD). For strong bursts, the receiver
may work beyond its linear range and a nonlinear calibration method
will be used instead (Yan et al., 2002).

In order to make our data more convincible, the other instruments
were also utilized to support the radio emission data. The soft
X-ray data from GOES was used to make a comparison. Also hard
X-ray observations with different energy ranges from RHESSI were
adopted. In addition, the photospheric magnetograph of the
line-of-sight magnetic component obtained from MDI/SOHO was
adopted to extrapolate and model the coronal magnetic field.

\subsection{Data Analysis}

Figure 1 presents the QPP event, which occurred at 06:24:15 --
06:24:21 UT, on 2005 July 30 in the frequency range of 2.60 -- 3.50
GHz. The upper and lower panels give the left- and right-handed
circular polarization, respectively. From this figure, we can see
that the QPP had a negative frequency drift rate (drift from high
frequency to the lower frequency) during 06:24:15 -- 06:24:18 UT
(named left wing, hereafter), and then the frequency drift rate
became positive (drift from low frequency to the higher frequency)
during 06:24:18 -- 06:24:21 UT (right wing), with the inflexion
occurring around 06:24:18 UT. The two yellow arrows indicate the
frequency drifting directions. With a linear fit we find that the
frequency drift rates at each wing of the QPP are -285 MHz
${s}^{-1}$ and 186 MHz ${s}^{-1}$, respectively.

In order to make sure that the QPP signals originate from the flare
bursts and they are not simply noise, Figure 2 presents three
profiles at the frequencies of 2.80, 3.00, and 3.20 GHz,
respectively. From the SGD database we may extrapolate that the
radio mean flux at frequency 2.60 -- 3.80 GHz of the quiet Sun on
2005 July 30 is about 100 -- 135 sfu. So the instrument sensitivity
is about $\triangle S/S_{\odot}\leq 2\% \simeq 2 - 2.7$ sfu. Figure
2 indicates that there are enhancements of more than 15 sfu in the
left and right wings of the QPP with respect to the background
emission. Moreover the enhancements around the QPP exceed the
instrument sensitivity greatly, so we may confirm that the QPP is
real, this dynamic spectrum is clear and reliable.

\begin{figure}
\begin{center}
 \includegraphics[width=10.0cm]{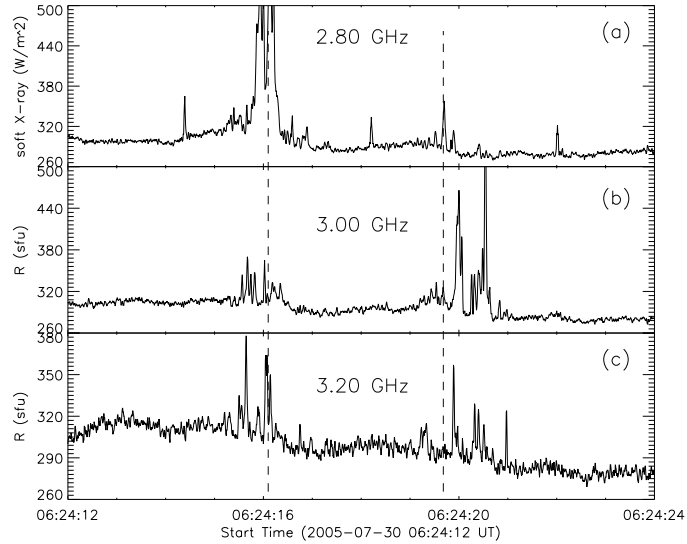}
  \caption{The profiles of radio emission at frequencies 2.80 GHz, 3.00 GHz and
3.20 GHz, respectively. The dashed lines mark the positions of the
maximum flux intensity at 2.80 GHz. The relative positions to the
dashed lines of the maximum flux intensity at 3.00 GHz and 3.20 GHz
reflect the frequency drift rates at the left and right part of the
R-DPS.}
\end{center}
\end{figure}

Figure 1 shows that the patterns or intensities of the QPP are
almost the same in the left-handed or right-handed polarization
spectrogram, indicating that the polarization of the QPP is not
obvious. Calculation indicates that the total polarization degree
($(R-L)/(R+L)$) is around -0.04\%, the polarization degree of left
wing is around -2.33\% and the right wing is around -3.22\%, where
R and L are the intensities of the right- and left- handed
circular polarization emission which subtract the background
components, respectively.

From the bright lines of the left and right wings of the structure
in Figure 1, we find that it is quasi-periodic, maybe it is hybrid
of more periodic components than one. The best way to analyze this
kind of structures is by using wavelet analysis, which can get
information on both the amplitude of any periodic component within
the series, and the temporal evolution of the QPP.

\begin{figure}
\begin{center}
 \includegraphics[width=10.0cm]{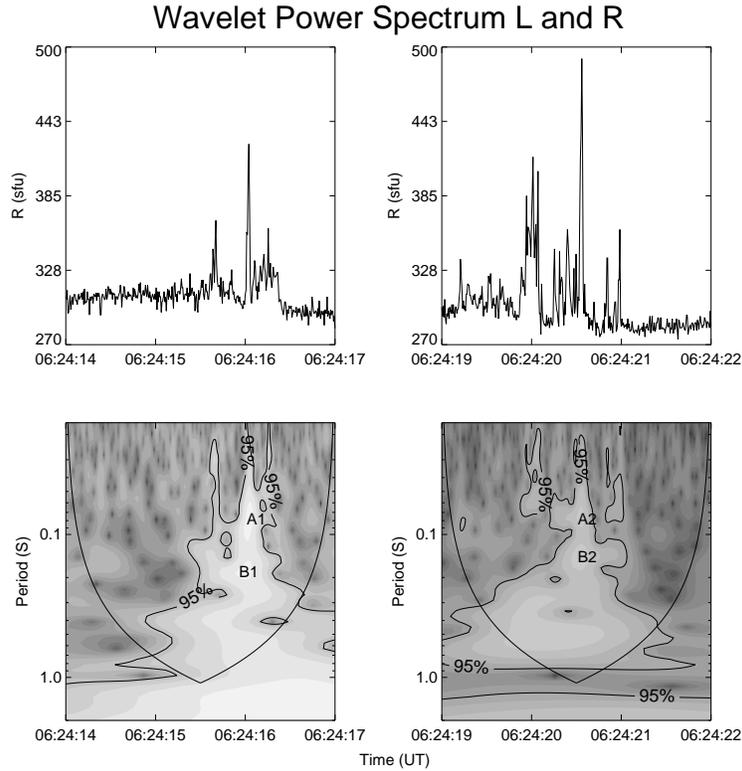}
  \caption{The bottom two panels show
the wavelet power spectrum at the left and right part of R-DPS,
using the Morlet wavelet. The black contours are the 95\% confidence
regions and anything "below" this line is dubious. The region below
the parabolic curve indicates the "cone of influence", where edges
influence is important. The A1 $\scriptsize{\sim}$ 80 ms, B1
$\scriptsize{\sim}$ 170 ms, A2 $\scriptsize{\sim}$ 70 ms, B2
$\scriptsize{\sim}$ 130 ms. The top two panels give corresponding
radio fluxes for time comparison.}
\end{center}
\end{figure}

Figure 3 presents the wavelet spectrum at frequency of 3.00 GHz
during 06:24:14 UT to 06:24:22 UT which just contains the time
interval of the QPP. The black contours plot the confident region
with 95\% confidence level. In the left part of the figure, there
are two obvious spectrum peaks corresponding to the left wing of
the QPP in the confident region, the periods are about 80 ms
(marked as A1) and 170 ms (marked as B1), respectively. This
implies that there are two pulsating components overlapped around
06:24:16 UT. On the right wing, the analogous structures appear
between 06:24:20 UT and 06:24:21 UT and the periods are about 70
ms (marked as A2) and 130 ms (marked as B2), respectively, which
are slightly shorter than that in the left wing. Both of them are
VSPs.

According to the analysis above, we could find some significant
relations between the left wing and right wings of this QPP.
Firstly, the gap between the two parts is only about 1 second,
which is much shorter than the duration of each part in the QPP;
secondly, the periods are very close in each part (80 ms at A1 to
70 ms at A2, 170 ms at B1 to 130 ms at B2, respectively) of the
emission frequency band; thirdly, both of the degrees of
polarization at each part of the QPP are not obvious. Therefore,
the name R-DPS should be more appropriate to describe such kind of
structures.

\begin{figure}
\begin{center}
 \includegraphics[width=10.0cm]{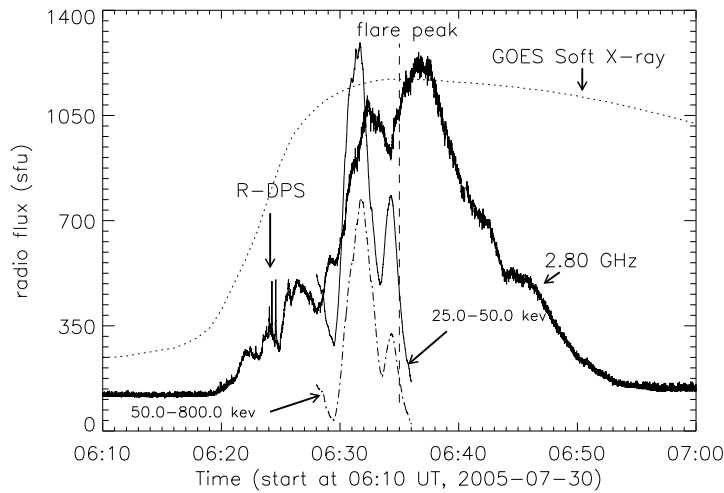}
  \caption{The relative position of the reversed drifting quasi-periodic pulsation (R-DPS) in
  the profile of the X1.3 flare event on July 30, 2005. The thick curve presents the profile of solar
  radio emission observed at frequency 2.80 GHz with SBRS/Huairou spectrometer. The thin curve
  and dot-dashed curve are the RHESSI hard X-ray curves in 25.0 -- 50.0 kev and 50.0 -- 800.0 kev, respectively. The dotted line
  shows soft X-rays from GOES satellite. The flare peak of this radio emission occurred at 06:35 UT.}
\end{center}
\end{figure}

Figure 4 presents context data to the microwave at 2.80 GHz (which
has a similar overall profile with that in the other frequencies),
GOES soft X-rays, and RHESSI hard X-rays associated with the X1.3
flare/CME event from 06:10 UT to 07:00 UT. Here, the R-DPS is marked
with a black arrow, which indicates that the R-DPS occurred in the
flare ascending phase, just after the onset of the flare. It is
associated with a time just when the gradient of the soft X-ray
reaches to its maximum. RHESSI also observed this flare event,
however, we have obtained RHESSI hard X-ray data only in the time
interval from 06:28 UT to 06:36 UT, and no valid data during the
R-DPS. Anyway, we still present the hard X-ray emission curves in 25
-- 50 keV and 50 -- 800 keV as reference.


The magnetic field configuration has importance for understanding
the physical processes of the QPP. Figure 5 gives the magnetic
topology of the flare active region AR 10792 obtained from
potential extrapolation computed from the observed line-of-sight
magnetic field using a Green's function. The initial version of
this technique is implemented by T. Metcalf and G. Barnes on 2005
October 25 and this program can be found in the SolarSoftWare
(SSW). The background of the magnetic extrapolation in the bottom
panel is the line-of-sight magnetogram observed by MDI/SOHO. Red
lines present closed magnetic field lines, while blue lines are
open field lines. Through the magnetic model, we can make rough
scale estimations of the coronal loops. However, as there is no
microwave imaging observation at the corresponding frequency, we
do not know which loop is associated with the R-DPS exactly. For
comparison, we present the H$\alpha$ image in the same area (the
upper panel in Figure 5). By reason of lacking data in the same
time range, here we just got a image of AR 10792 at 06:35:49 UT on
2005 July 30, observed by BBSO, while it is still valuable for
comparison with the extrapolated model. It is obvious that there
is a two-ribbon flare in this image. This structure often
indicates that magnetic reconnection has allowed the coronal
magnetic field to relax into a lower energy state. Practically, it
is natural to assume that only the coronal loops which are
adjacent to the flare ribbons are related to the microwave bursts.
From Figure 5 we may obtain the lengths of these coronal loops are
about 2" -- 100". Suppose the coronal loops are semicircles, then
the lengths of the coronal loops are about $2.3\times10^3$ --
$1.14\times10^5$ km.

\begin{figure}
\begin{center}
 \includegraphics[width=10.0cm]{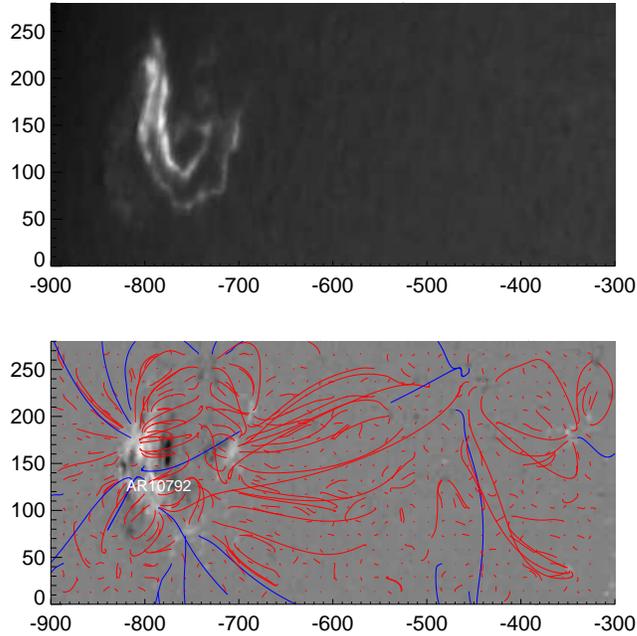}
  \caption{Potential extrapolation of magnetic field lines in the lower panel
using the observed line-of-sight field of the MDI/SOHO during
06:24:12 -- 06:24:24 UT, on July 30, 2005. Coordinates are in
arcseconds. The coordinate of the center of the Sun is (0,0). The
image in the upper panel is from the H$\alpha$ data of the Big Bear
Solar Observatory (BBSO) at 06:35:49 UT. There is an obvious
two-ribbon flare in this image.}
\end{center}
\end{figure}

\section{Discussion on the possible process of R-DPS}

According to the work of Tan et al. (2007), Tan (2008) and Tan et
al. (2010), VSP can be explained as a result of modulations of the
resistive tearing-mode oscillation in some electric current-carrying
flare loops. The pulsating emission is possibly plasma emission. As
we know that the plasma emission is always generated at the plasma
fundamental frequency ($\omega_{pe}$) or at the second harmonic
frequency ($\sim2\omega_{pe}$). The degree of polarization of
fundamental plasma emission is very strong and usually in the sense
of O-mode, while the second harmonic plasma emission is always a
weak circular polarization. As the R-DPS is weakly polarized, we may
suppose that it is possibly related to the second harmonic plasma
emission. The central frequency of the R-DPS is about 3.00 GHz, and
implies that the plasma density is about $2.78\times
10^{10}$cm$^{-3}$. Plasma with such high density is probably very
close to the flare core.

Based on the plasma emission mechanism, we have the emission
frequency: $f=sf_{pe}\simeq9sn_{e}^{1/2}$, we may obtain the
frequency drift rate as:

\begin{equation}
\frac{df}{dt}\simeq\frac{f}{2H}v
\end{equation}

Here, $H=\mid n_{e}/\frac{dn_{e}}{dr}\mid$ is the inhomogeneous
scale length of the plasma in the source, $v=\frac{dr}{dt}$ is the
moving velocity of the emission source region. Then we may get the
moving velocity: $v=\frac{2H}{f}\frac{df}{dt}=2H\varepsilon$,
$\varepsilon=\frac{1}{f}\frac{df}{dt}$ is the relative frequency
drift rate. From here we know that the moving velocity is only
proportional to the relative frequency drift rate. Usually, the
inhomogeneous scale length $H$ should be induced from the solar
active region atmospheric model. For simplicity, we may assume
that $H\sim10^{4}$ km. Then we may estimate that the source moving
velocity associated with the left wing of the R-DPS is about 1900
km s$^{-1}$ , and 1240 km s$^{-1}$ in opposite direction with the
right wing. This may imply that the R-DPS reflects a following
process: during the rising phase of the X1.3 flare, the closed
flaring coronal loop has an upthrust in velocity of 1900 km
s$^{-1}$, and then falls down slowly in velocity of 1240 km
s$^{-1}$.

To the explanation of the loop upward and downward movings, a
two-dimensional (2D) resistive-MHD numerical simulation of the
reconnection starting from the Harris-type current sheet has been
done (B\'arta, V\v{s}nak, and Karlick\'y, 2008). The result of
simulation indicated that if the reconnection rate v$\times$B at the
X-point below the plasmoid is higher than the one at the X-point
above the plasmoid, the plasmoid moves upward since the net tension
causes an upward electron acceleration and then excites the plasma
emission in the upper source region. If the magnetic flux is
reconnected in the upper diffusion region is higher than in the
lower one, the plasmoid moves downward and the high energy electron
flow excites plasma emission from the lower source region. However
as it is stated above, if the source regions are located at
different altitudes, the density of the source regions would be very
different, depending on the altitudes, and then different waveband
signals from the right and left wing of the R-DPS would be received.
This does not agree with our observation, that the frequency range
of the R-DPS event from 2.60 GHz to 3.80 GHz. Therefore we take
another way to interpret our observations.

If the emission source region could be located within a loop with
up-and-down motions, it would be more consistent with the
observation. We assume that the up and down motions corresponded to
the expansion and shrinkage of the loops. These processes should
have a relation with intense energy injection (Li and Gan, 2005).
During the shrinkage of the loops, there were few intense energy
injections, since the chromospheric evaporation needed several
minutes to fill the loops, and during this time the density of loops
was rather low while it was opposite around the loops. Afterward,
the injection process completed, the density of the region above the
loop top was lower corresponding to the loop system, then the loop
began to expand.

We may assume that the flaring loop is current-carrying plasma loop,
having an up-and-down motion, a resistive tearing-mode instability
will be triggered in the flaring loop and a series of multi-scale
magnetic islands would form. Electron acceleration will occur at
X-points between every two adjacent magnetic islands. Then the
energetic electrons will excite some Langmuir turbulence in the
flare plasma loop and make the plasma emission enhanced. Modulated
by the resistive tearing-mode oscillation, the emission will behave
as pulsating structure in the spectrogram.

At the same time, we know there are two pulsating components in both
the left and right wings of the R-DPS, and this may indicate that
there are two different flaring plasma loops in the same oscillating
source region. The difference may be in loop radius, or electric
current, etc. (Tan, 2008). However, as we have no corresponding
imaging observations, we could not confirm which factor is the real
candidate.

\section{Conclusions}

In this work we present detailed observations of a particular
reversed drifting quasi-periodic pulsation (R-DPS) associated with
the rising phase of an X1.3 flare event. From the above data
analysis and discussions, we may reach the following conclusions:

(1) It is observationally confirmed that the theoretically predicted
reversed direction frequency drift structures in microwave emission
indeed exist.

(2) The R-DPS is mainly composed of two pulsating components: one is
a slow-VSP with period of about 130 -- 170 ms, the other is a
fast-VSP with period of about 70 -- 80 ms.

(3) The frequency drift rate in the left wing of the R-DPS is about
-285 MHz ${s}^{-1}$, and in the right wing about 186 MHz ${s}^{-1}$.

(4) The polarization of the R-DPS is a weak left-handed circular
polarization.

Based on the assumption of plasma emission mechanism that the
tearing mode oscillation modulates the plasma emission in
current-carrying plasma loops, the R-DPS may reflect the dynamic
processes of the emission source regions. From the frequency drift
rates we make an estimation of the source up-and-down motion
velocity being about 1900 km s$^{-1}$ up and then 1240 km s$^{-1}$
down. The variations of the plasma density in the loop with respect
to the background during the up-and-down motion result the reversed
drifting quasi-periodic pulsations. In order to confirm this
deduction, some microwave imaging observations at the corresponding
frequency is necessary. The constructing {\it Chinese Spectral
Radioheliograph }(0.4 -- 15 GHz) will satisfy this need (Yan et al,
2009).

\begin{acknowledgements}

The authors would like to thank the referee for the helpful and
valuable comments on this paper. We would also thank the the GOES,
RHESSI, MDI/SOHO, BBSO and SBRS/Huairou teams for providing
observation data. This work was supported by NSFC Grant No.
10873021, 10921303, 10903013, 11103044, 11103039, MOST Grant No.
2011CB811401, and the National Major Scientific Equipment R\&D
Project ZDYZ2009-3.

\end{acknowledgements}

\end{article}
\end{document}